\documentclass[aps,twocolumn,showpacs,preprintnumbers,nofootinbib,prd,superscriptaddress,groupedaddress,10pt]{revtex4-1}

\makeatletter
\def\l@subsubsection#1#2{}
\def\l@subsubsub section#1#2{}
\makeatother

\usepackage{amsmath}
\usepackage{graphicx,amssymb,amsmath,amsthm,amsfonts,epsfig,epsf}
\usepackage[usenames,dvipsnames,svgnames,table]{xcolor}
\usepackage{epstopdf}
\definecolor{darkred}{rgb}{0.5,0,0}
\usepackage{aas_macros}
\usepackage{bm}
\usepackage{dcolumn}
\usepackage[utf8]{inputenc}
\usepackage{latexsym}
\usepackage{rotating}
\usepackage{longtable}

\usepackage{enumerate}
\usepackage{tensor}
\usepackage{multirow}
\usepackage{booktabs}
\usepackage{mathtools}
\usepackage{url}
\usepackage[linktocpage,hidelinks]{hyperref}

\newcommand{\oo}{\left(}
\newcommand{\cc}{\right)}
\newcommand{\ooq}{\left[}
\newcommand{\ccq}{\right]}
\newcommand{\mndd}{_{\mu\nu}}

\newcommand{\ijdd}{_{ij}}
\newcommand{\ijuu}{^{ij}}

\newcommand{\rgb}{\mathcal{R}_{\rm GB}}
\newcommand{\til}{~}

\usepackage{colortbl}

\def\nn{\nonumber}

\def\be{\begin{equation}}
\def\ee{\end{equation}}
\newcommand{\beq}{\begin{eqnarray}}
\newcommand{\eeq}{\end{eqnarray}}

\newcommand{\nocontentsline}[3]{}
\newcommand{\tocless}[2]{\bgroup\let\addcontentsline=\nocontentsline#1{#2}\egroup}
\def\ba{\begin{align}}
\def\ea{\end{align}}

\newcommand{\warn}[1]{{\textcolor{red}{\sf{[IN PROGRESS]}} }}

\begin{document}

\title{CLAP for modified gravity:
scalar instabilities in binary black hole spacetimes}

\author{
Lorenzo Annulli
}

\affiliation{Centro de Astrof\'{\i}sica e Gravita\c c\~ao  - CENTRA, Departamento de F\'{\i}sica, Instituto Superior T\'ecnico - IST, Universidade de Lisboa - UL, Av. Rovisco Pais 1, 1049-001 Lisboa, Portugal}
%

\begin{abstract}
The close limit approximation of binary black hole is a powerful method to study gravitational-wave emission from highly non-linear geometries. In this work, we use it as a tool to model black hole spacetimes in theories of gravity with a new fundamental scalar degree of freedom. As an example, we consider Einstein-scalar-Gauss-Bonnet gravity, which admits as solution the Schwarzschild geometry
as well as black holes with scalar hair. Accordingly, we find scalar perturbations growing unbounded around binary systems. This ``dynamical scalarization'' process is easier to trigger (i.e. occurs at lower values of the coupling constant of the theory) than the corresponding process for isolated black holes. Our results and framework highlight the fundamental role of the interaction during the collision of compact objects. They also emphasize the importance of having waveforms for black hole binaries in alternative theories, in order to consistently perform tests beyond General Relativity.
\end{abstract}

\maketitle

\noindent{\bf{\em I. Introduction.}}
The LIGO/Virgo detections of gravitational waves (GWs) produced by coalescing black holes (BHs) and stars provided the first insights on regimes where dynamical gravitational interactions dominate over the other known fundamental forces (also known as the {\it strong gravity regime})~\cite{Abbott:2016blz,Abbott:2016nmj,Abbott:2017vtc,Abbott:2017oio,Abbott:2020tfl}. These events 
provided important constraints on the general relativistic theory of gravitation (GR), and on some modified gravity models~\cite{TheLIGOScientific:2016src,Barack:2018yly,
Bird:2016dcv,Cornish:2017jml,Ezquiaga:2017ekz,Creminelli:2017sry,Annala:2017llu,Yunes:2016jcc,Baker:2017hug,Sakstein:2017xjx,Cardoso:2019rvt}. Despite the very good agreement between GR predictions and the observed signals, there are still fundamental phenomena that GR is not able to explain thoroughly. For instance, the lack of a profound understanding of the nature of singularities~\cite{Penrose:1964wq,Penrose:1969,Cardoso:2017soq,Cardoso:2017cqb,Cardoso:2019rvt}, or the origin of dark energy or dark matter~\cite{Weinberg:1988cp,Bertone:2018krk} show that there is still room for possible extensions or modifications of Einstein's theory. 

The advent of third generation detectors~\cite{Hild:2010id,Punturo:2010zza,Maggiore:2019uih} and the space-based LISA mission~\cite{Audley:2017drz} will increase the number and accuracy of GW observations, paving the way to a new, {\it precision gravitational wave astronomy} era. Data from massive and distant compact objects will provide a statistical and systematic vision of the objects populating our Universe. Precision studies will help in assessing foundational questions about the ultimate nature of the gravitational theory itself. In fact, tests of GR and its alternatives are based on the capability to constrain the parameters of each theory with the highest precision. 

Tests of gravity comprise also smoking-guns for new physics. These unique predictions of an alternative theory, may therefore allow to discriminate between GR and its competitors. In view of this, it is crucial to search for such peculiar mechanisms in the GW signals produced by compact bodies\til\cite{Damour:1993hw,Cardoso:2011xi,Berti:2015itd,Barausse:2020rsu,Brito:2015oca}. A representative example of such phenomena occurs in the framework of scalar-tensor theories, for example, where a new fundamental scalar degree of freedom couples to matter with some strength $\beta$. For certain coupling strengths $\beta$ one finds static solutions in scalar tensor theory with a trivial scalar, equivalent to those of GR, and which are stable solutions. However, there are couplings for which a GR solution is unstable and triggers a ``tachyonic'' instability, leading to stars or BHs with nontrivial charge~\cite{Damour:1992we}. These bodies are said to be {\it scalarized}~\cite{Ruffini:1971bza,Hawking:1972qk,Sotiriou:2011dz,Sotiriou:2013qea,Sotiriou:2014pfa,Silva:2017uqg,Doneva:2017bvd,Witek:2018dmd,Silva:2018qhn,Minamitsuji:2018xde,Doneva:2019vuh,Fernandes:2019rez,Minamitsuji:2019iwp,Cunha:2019dwb,Andreou:2019ikc,Ikeda:2019okp}. The possibility to ``awake'' a new fundamental field is a valuable smoking gun for these alternative theories, as it leads to dipolar emission of radiation for example. Due to its non-perturbative nature, spontaneous scalarization of neutron stars avoids the strong constraints set by solar system experiments, established in the regime where the gravitational forces are relatively close to the Newtonian ones\til\cite{Will:2014kxa,Ramazanoglu:2016kul}. Additionally, scalarization phenomena may occur also for vector, tensor and spinor fields\til\cite{Ramazanoglu:2017xbl,Doneva:2017duq,Annulli:2019fzq,Kase:2020yhw,Ramazanoglu:2019gbz,Ramazanoglu:2017yun,Ramazanoglu:2018hwk,Minamitsuji:2020hpl}.
%
%

Recent work on the scalarization of multi-body systems showed how signatures of this non-perturbative mechanism can emerge dynamically\til\cite{Barausse:2012da,Palenzuela:2013hsa,Shibata:2013pra,Silva:2020omi,Cardoso:2020cwo}. The main objective of this work is to study this dynamical scalarization process in binary BH (BBH) spacetimes, in theories allowing for spontaneous scalarization of isolated BHs. In other words, working with non-trivial couplings between the scalar and the spacetime curvature, we wish to highlight the effects of scalar field dynamics in a two-body spacetime. In the following we consider Einstein-scalar-Gauss-Bonnet gravity (EsGB) as a specific example of scalar-tensor theory of the above class. EsGB emerges naturally in the low-energy limit of string theories~\cite{METSAEV1987385,Kanti:1995vq,Charmousis:2014mia}, and is the only alternative theory that includes an extra scalar degree-of-freedom, coupled to a quadratic curvature term constructed from the spacetime metric, which equations of motion are second (differential) order. In order to model BBH configurations, we use results from the Close Limit Approximation (CLAP) of binary BHs\til\cite{Price:1994pm,Anninos:1995vf,Abrahams:1995wd,Andrade:1996pc}. This perturbative method was used to find the ringdown waveforms produced by the head-on collision of BHs binaries in GR, and it was recently generalized to less standard scenarios\til\cite{Annulli:2021dkw}.
%

%

Units are such that $G=c=\hbar=1$.

\noindent{\bf{\em II. Einstein-scalar-Gauss-Bonnet gravity.}}
To study GW generation in modified gravity, a thorough study of the properties of the theory is needed. Namely, carrying out a spacetime decomposition (e.g. 3+1)\til\cite{Arnowitt:1962hi,Gourgoulhon:2007ue,alcubierre2008introduction,baumgarte2010numerical,shibata2015numerical}, understanding if the theory is well-posed, constructing physically motivated initial data and performing their time evolution. This program has been carried out for only a few theories~\cite{Salgado:2005hx,Salgado:2008xh,Berti:2013gfa,Shibata:2013pra,Torii:2008ru,Yoshino:2011qp}. For the above-mentioned EsGB theory, a 3+1 decomposition of the field equations has been recently performed\til\cite{Witek:2020uzz,Julie:2020vov}.

The action of EsGB is given by
\be \label{eq:EsGB_action}
S=\frac{1}{16\pi}\int d^4x\sqrt{-g}\left[R-\frac{1}{2}\left(\nabla \Phi\right)^2+\frac{\eta}{4} \mathfrak{f}(\Phi)\rgb\right]\,,
\ee
where $\eta$ is the dimensionful coupling constant of the theory and $f\oo\Phi\cc$ is a generic coupling function between the scalar field and the Gauss-Bonnet invariant $\rgb$, that is
\be
\rgb=R^2-4R\ijdd R\ijuu +R_{ijkl}R^{ijkl}\,,
\ee
with $R\oo R_{ij}\cc$ being the Ricci scalar (tensor) and $R_{ijkl}$ the Riemann tensor. The equations of motion corresponding to the action\til\eqref{eq:EsGB_action} are given by
\beq
\label{eq:EsGB_einsteineq}
G_{\mu\nu}&=&\frac{1}{2} T\mndd-\frac{1}{8}\eta\,\mathcal{G}\mndd\,,\\
\label{eq:EsGB_scalarKG}
\square\Phi&=&-\frac{\eta}{4} \frac{\partial \mathfrak{f}\oo\Phi\cc}{\partial\Phi}\rgb\,,
\eeq
where $G_{\mu\nu}$ is the usual Einstein tensor and
\beq
\mathcal{G}\mndd&=& 16 R^\alpha_{(\mu}\mathcal{C}_{\nu)\beta} +8\mathcal{C}^{\alpha\beta}\oo R_{\mu\alpha\nu\beta}-g\mndd R_{\alpha\beta} \cc \nn\\
&&-8 \mathcal{C} G\mndd -4 R \mathcal{C}\mndd\,,
\eeq
with 
\be
\mathcal{C}\mndd= \nabla_\mu \nabla_\nu \mathfrak{f}\oo\Phi\cc=\mathfrak{f}'\nabla_\mu\nabla_\nu\Phi+\mathfrak{f}''\nabla_\mu\Phi\nabla_\nu\Phi\,,
\ee
and the scalar field stress-energy tensor is defined as,
\beq
T_{\mu\nu}&=&\partial_\mu\Phi\partial_\nu\Phi-\frac{1}{2}g\mndd \partial^\alpha\Phi\partial_\alpha\Phi\,.
\eeq
%

In order to study the evolution of any physical configuration in EsGB, one needs to find consistent initial data. This consists in solving the EsGB constraint equations coming directly from Eqs.\til\eqref{eq:EsGB_einsteineq}-\eqref{eq:EsGB_scalarKG}.
A solution to these equations, in general, includes complicated functions of the scalar $\Phi$ and the scalar momentum density $K_{\Phi}$\footnote{$K_{\Phi}$ is defined as the Lie derivative of the scalar field with respect to the normal vector to the initial hypersurface of foliation.}. However, in this work we are only interested in understanding if, and how, BBHs in vacuum might be unstable in EsGB. In order to assume a trivial scalar field profile $\Phi=0$ and momentum density $K_{\Phi}=0$ on the initial hypersurface, we restrict to theories obeying $d \mathfrak{f}/d \Phi\rvert_{\Phi=0}=0$. With this assumption, we rule out theories allowing only for BH solutions with scalar hair, as the ones due to an exponential coupling function (see Ref.\til\cite{Kanti:1995vq}). Further considering BHs initially at rest, the momentum constraint equations are identically satisfied (and therefore not shown here), while the Hamiltonian reads as in vacuum GR
\be \label{eq:GR_reduced_ID_eq}
\prescript{3}{}{R}=0\,,
\ee
where $\prescript{3}{}{R}$ is the Ricci scalar evaluated on the initial three-spacelike hypersurface of foliation.

\noindent{\bf{\em III. Binary black hole spacetime.}}
In order to model a BBH spacetime, one needs to account for their interaction energy. The CLAP formalism of BBHs in GR succeeded to consistently describe such configurations\til\cite{Price:1994pm,Anninos:1995vf,Abrahams:1995wd,Andrade:1996pc,Annulli:2021dkw}, and we will use this approximation in what follows.
This approach is based on having initial data describing BBHs that are solutions of the Hamiltonian constraint equation\til\eqref{eq:GR_reduced_ID_eq}. Such solution is not unique: different initial data\til\cite{Misner:1960zz,Brill:1963yv,Bowen:1980yu} may be used within the CLAP. The ones that we employ in this work are given by the Brill-Lindquist (BL) initial data\til\cite{Brill:1963yv}. These are conformally flat, time symmetric initial data representing two BHs initially at rest. 

Let us focus on equal-mass binaries, of total ADM mass $M$. In isotropic cartesian coordinates, we place the BHs on the $Z$-axis (${\bm R_{1/2}}=(0,0,\pm Z_0)$, where ${\bm R_i}$ is the position of each BH in this reference), therefore the origin of the reference frame is in the center-of-mass of the system. As shown in detail in Refs.\til\cite{Abrahams:1995wd,Andrade:1996pc,Sopuerta:2006wj,Annulli:2021dkw}, using the CLAP of BBHs, we can recast the $4D$ initial spacetime as a perturbation of the Schwarzschild metric. Thus, including for the sake of simplicity only the leading-order quadrupolar contribution\til\cite{Price:1994pm}, the spacetime can be written as 
\begin{equation}\label{eq:pert_Schw_CLAP_GR_BL}
g_{\mu\nu}=g^{(0)}_{\mu\nu}+h_{\mu\nu}\,,
\end{equation}
where 
\begin{equation}
g^{(0)}_{\mu\nu}={\rm
  diag}(-f,f^{-1},r^2,r^2\sin^2\theta)\,,
\end{equation}
and, using the Legendre polynomial $P_2 \left(\cos\theta\right)$, $h_{\mu\nu}$ is given by 
\begin{align}
h_{rr}&=f^{-1}gP_2(\cos\theta)\frac{Z_0^2}{2M^2}\,,\nonumber\\
h_{\theta\theta}&=r^2g P_2(\cos\theta)\frac{Z_0^2}{2M^2}\,,\label{eq:BLrecast1}
\end{align}
with 
\begin{equation}
g=4\left(1+M/(2R)\right)^{-1}M^3/R^3\,,
\end{equation}
and the isotropic coordinate $R$ is defined in terms of the Schwarzschild radial coordinate $r$ as
\begin{equation}
\label{eq:R_isotropic_to_Schw}
R=\frac{1}{4}\left(\sqrt{r}+\sqrt{r-2M}\right)^2\,.
\end{equation}
%

The parameter $Z_0$ in Eq.~\eqref{eq:BLrecast1} represents the initial separation between the BHs in the isotropic frame. For $Z_0=0$ there is just a single BH of mass $M$ in the initial slice. When $0<Z_0\lesssim 0.4$ one single common horizon appears\til\cite{Anninos:1995vf,Abrahams:1995wd,Andrade:1996pc}. In this regime, the spacetime can be thought to represent two BHs close to one another, enveloped by a common distorted horizon~\cite{Price:1994pm}. Moreover, it is worth to note that $Z_0$ itself is only a parameter and not a physical quantity. However, it is possible to establish a realation between $Z_0$ and the physical distance between the apparent horizons of the initial colliding BHs ($L$)~\cite{Andrade:1996pc,Annulli:2021dkw,bishop1982closed,bishop1984horizons,Gleiser:1998rw,Sopuerta:2006wj}: an explicit computation gives $L=3M$ for $Z_0\simeq0.5M$, $L=3.5M$ for $Z_0\simeq0.7M$, $L=4M$ for $Z_0\simeq0.85M$.

The metric in Eq.\til\eqref{eq:pert_Schw_CLAP_GR_BL} shows how the colliding BHs spacetime can be seen as a time-dependent perturbation of a Schwarzchild background. Hence, in the CLAP, the time evolution of this small (even) gravitational perturbations can be achieved by gauge-invariant perturbations techniques~\cite{Moncrief:1974am,Cunningham:1978zfa,Cunningham:1979px}. Notably, as shown in Ref.\til\cite{Price:1994pm}, the gravitational perturbation equations can be cast in a single Zerilli equation for one unknown function (the Zerilli function)\til\cite{Zerilli:1970se}. Solutions of such equation provide GW signals remarkably similar to the results obtained using full numerical simulations\til\cite{Anninos:1993zj}.

Instead, in the following, we use the metric in Eq.\til\eqref{eq:pert_Schw_CLAP_GR_BL} only as the background spacetime in which evolving the scalar field, thus neglecting the motion of the BHs in the timescale of the oscillation. This is a severe approximation. First because astrophysical BHs in binaries move at large velocities when close to one another. Furthermore, on a timescale of order $M$, the BHs collide, hence the extrinsic spacetime curvature will take non-zero values, changing the background spacetime in which scalar perturbations propagate. However, albeit an approximation, restricting to a {\it frozen} background still shows the main feature of the onset of instabilities in binary spacetimes, as we shall see later.
 
A CLAP treatment allowing for spontaneous scalarization {\it during} the collision (or the inspiral) of BHs in EsGB is left for future work.

\noindent{\bf{\em IV. Scalar instabilities.}}
To test the onset of scalar instabilities in BBHs geometries, we study the behaviour of small linear scalar fluctuations in backgrounds described by Eq.\til\eqref{eq:pert_Schw_CLAP_GR_BL}. These vacuum configurations have been chosen since EsGB allows also for BH solutions identical to GR.


Small scalar perturbations can be mathematically expressed replacing $\Phi \rightarrow \epsilon\Phi$ in Eqs.\til\eqref{eq:EsGB_einsteineq}-\eqref{eq:EsGB_scalarKG}, with $\epsilon$ a small bookkeeping parameter. Thus, one can linearize the Einstein-KG system up to $\mathcal{O}(\epsilon)$. In this limit, the KG equation decouples from Einstein's equations. Hence, the background spacetime is not affected by the scalar perturbations. Our perturbation scheme will eventually breakdown at sufficiently late times: the exponentially growing scalar gives rise to an exponentially growing stress-tensor, the backreaction of which on the geometry can no longer be neglected. Here, we focus solely on the early-time development of the instability. 

What we are left to solve is the KG equation
\be \label{eq:KG_equation}
\square\Phi=-\frac{\eta}{4} \frac{\partial \mathfrak{f}\oo\Phi\cc}{\partial\Phi}\rgb\,,
\ee
where the box operator ($\square=\frac{1}{\sqrt{-g}}\partial_\mu\oo g^{\mu\nu}\sqrt{-g}\partial_\nu\cc$) is defined on the BBH background in Eq.\til\eqref{eq:pert_Schw_CLAP_GR_BL}. Let us further assume a quadratic Gauss-Bonnet coupling function
\be \label{eq:coupling_function}
\mathfrak{f}\oo\Phi\cc=\frac{\Phi^2}{2} \,.
\ee
As shown in Refs.\til\cite{Silva:2017uqg,Doneva:2017bvd}, in this class of theories the KG equation admits solutions composed by a constant scalar around spacetimes satisfying GR equations. Furthermore, a linear stability analysis showed that, for certain values of the coupling constant $\eta$, GR solutions may be unstable. To find the endpoint of this instability, one needs to solve the equation of motion including the backreaction of the scalar on Einstein's equations. This eventually leads to scalarized (or hairy) BHs or stars.

Conversely, here we are interested in the effect on scalar fluctuations due to the presence of a binary. Hence, both the box operator and $\mathcal{R}_{\rm GB}$ in Eq.\til\eqref{eq:EsGB_scalarKG} depend on the perturbed BBH spacetime, and in the CLAP, we may expand them in powers of the small BHs separation $Z_0$,
\beq \label{eq:box_expansion}
\square &=& \square^{(0)}+ Z_0^2 \, \square^{(1)} + \mathcal{O}(Z_0^3)\,,\nn\\
\rgb &=& \rgb^{(0)}+ Z_0^2 \, \rgb^{(1)} + \mathcal{O}(Z_0^3)\,.
\eeq
Decomposing the scalar in spherical harmonics as,
\be \label{eq:Phi_ansatz}
\Phi\oo t,r,\theta,\varphi\cc=\frac{1}{r}\sum_{\ell ,m}\psi_{\ell m}\oo t,r\cc Y^{\ell m}\oo\theta,\varphi\cc\,,
\ee
the KG equation is non-separable because it couples different components of the index $\ell$. The mathematical details of the procedure to separate (perturbatively) Eq.\til\eqref{eq:EsGB_scalarKG} are given in  appendix\til\ref{app:separateKG}. Let us summarize here the most important passages to arrive to the master equation that describes scalar perturbations in EsGB, in the axisymmetric stationary BBH spacetime. 

The procedure is similar to the one described in Ref.\til\cite{Annulli:2021dkw,Cano:2020cao}. Let us expand the scalar field using the spherical harmonics base. The key point is that, for each spherical harmonics index $\ell$, the KG equation becomes separable in the limit $Z_0 \rightarrow 0$. Hence, using the specific ansatz in Eq.\til\eqref{eq:Phi_ansatz_2}, for each $\ell\geq 1$\footnote{The monopolar $\ell =0$ perturbations are not affected by the $Z_0^2$ corrections (see Eq.\til\eqref{eq:q12lm}), hence the $\ell =0$ modes are the same as in the single BH case.} one gets a Schr\"{o}dinger-like equation that includes corrections in $Z_0^2$,
\beq \label{eq:KG_notortoise}
&&\frac{\partial^2 \psi_{\ell m}}{\partial t^2}+\frac{\partial^2 \psi_{\ell m}}{\partial r^2} \oo U_0+Z_0^2\tilde{U}_0\cc+\frac{\partial \psi_{\ell m}}{\partial r}  \oo U_1+Z_0^2\tilde{U}_1\cc\nn\\
&&+\psi_{\ell m} \oo \oo W_0+\frac{\eta}{4}\tilde{W}_0\cc +Z_0^2 \oo W_1+ \frac{\eta}{4} \tilde{W}_1\cc\cc=0\,,\nn\\
&&
\eeq
where all the potentials are listed in Eq.\til\eqref{eq:potentials_notortoise}. Setting $\eta/M^2=Z_0/M=0$ in Eq.\til\eqref{eq:KG_notortoise}, one gets the perturbations describing scalar perturbations in a static Schwarzschild spacetime\til\cite{Berti:2009kk}. Additionally, one may notice that the scalar field fluctuations are independently affected both by $\eta$ and $Z_0$. This means that there might be non-trivial effects on the scalar quasi normal modes of oscillation of a BBH even in pure GR (setting $\eta=0$ and $Z_0\neq 0$). For such scenario, we refer the interested reader to Ref.\til\cite{Annulli:2021dkw}. In the following we strictly focus on EsGB (hence $\eta \neq0$).

\noindent{\bf{\em Boundary conditions.}}
%
To find unstable modes, we start with an harmonic time dependent scalar field, 
\be \label{eq:phi_harmonic_timedep}
\psi_{\ell m}\oo t,r\cc=\Psi\oo \omega,r\cc e^{-i \omega t}\,,
\ee
where we dropped the subscript $_{\ell m}$ in the r.h.s.. Substituting the ansatz\til\eqref{eq:phi_harmonic_timedep} in Eq.\til\eqref{eq:KG_notortoise}, an unstable mode is found when a bounded regular solution of the KG equation
\beq \label{eq:KG_BBH}
&&\frac{\partial^2 \Psi}{\partial r^2}  \oo U_0+Z_0^2\tilde{U}_0\cc+\frac{\partial \Psi}{\partial r}\oo U_1+Z_0^2\tilde{U}_1\cc\nn\\
&&+\Psi \oo \oo W_0+ \frac{\eta}{4}\tilde{W}_0\cc+Z_0^2 \oo W_1+ \frac{\eta}{4} W_2\cc -\omega^2\cc=0\,,\nn\\
&&
\eeq
with potentials in Eq.\til\eqref{eq:potentials_notortoise}, possesses a frequency that satisfies 
\be
\omega=\omega_R+i\omega_I,\;\;\text{with }\omega_I>0	.
\ee
%
%
Being interested in the onset of the instability, without loss of generality, we might look for solutions with purely imaginary frequencies ($\omega_R=0$). The asymptotic behaviours of Eq.\til\eqref{eq:KG_BBH} provide us  the proper boundary conditions to be imposed. Especially, asking for regularity both at the horizon and at spatial infinity, we get %
\beq \label{eq:BC_boundstates}
&&\Psi \oo r \sim 2M\cc = 
\oo r-2M\cc^{\frac{2  M  \omega_I}{\sqrt{1-2 \oo Z_0/M\cc^2 q^{(1)}_{\ell m}} }} \sum_{n=0}^N a_n \oo r-2M\cc^n\,,\nn\\
&&\Psi \oo r \sim\infty\cc = \frac{e^{- r \omega_I}}{r^l}\sum_{n=0}^N b_n r^{-n}\,,
\eeq
where the coefficients $a_n,b_n$ have to be found substituting Eq.\til\eqref{eq:BC_boundstates} in  Eq.\til\eqref{eq:KG_BBH} and solving it order by order. For each configuration, the value of $N$ has to be increased until the boundary conditions\til\eqref{eq:BC_boundstates} do not converge to fixed values\til\cite{Pani:2013pma}.

\noindent{\bf{\em Isolated black hole scalar bound states.}}
\begin{figure}[ht]
\includegraphics[width=7.5cm,keepaspectratio]{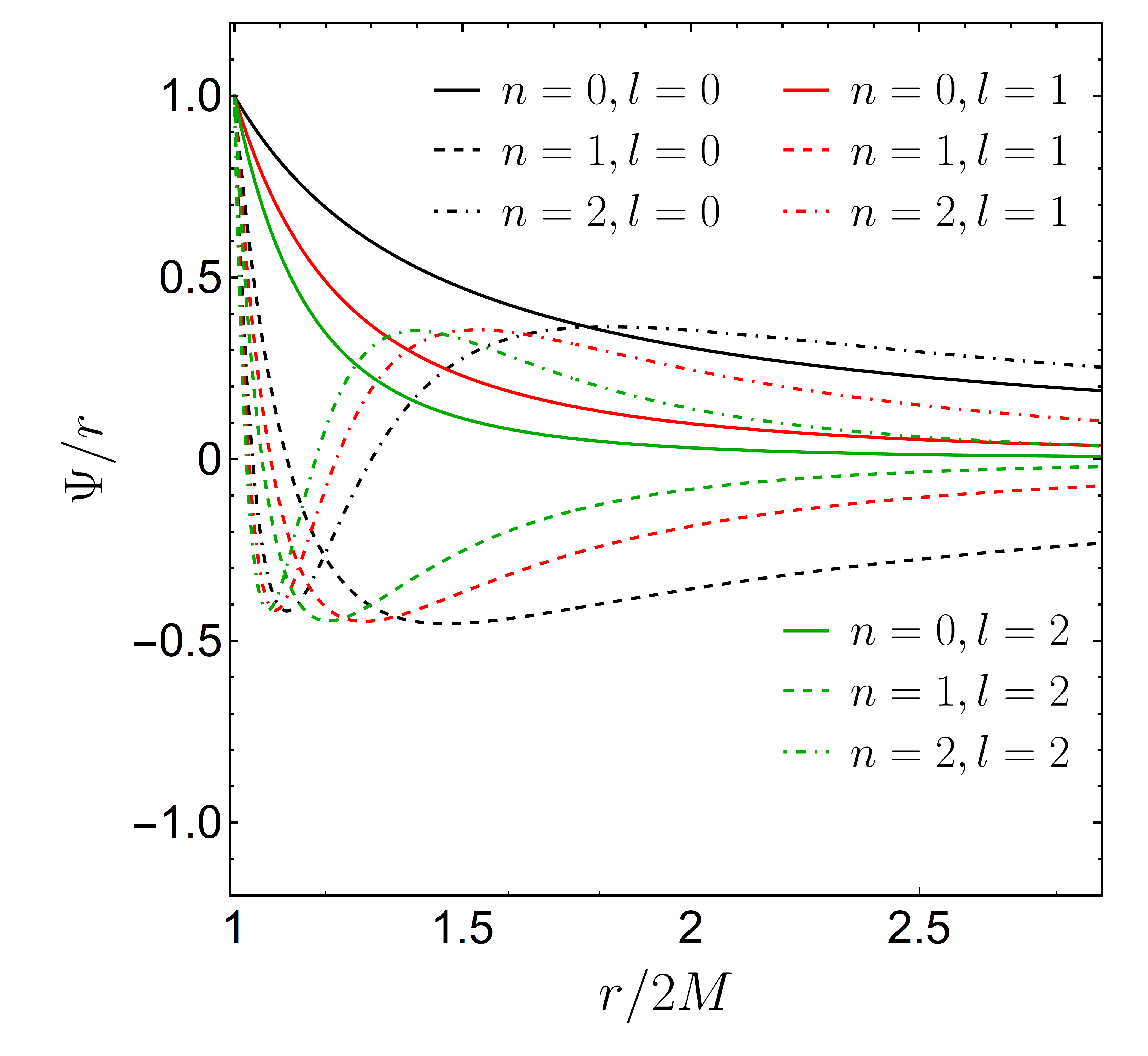} 
\caption{Scalar profiles for different values of $\ell$, for the first three scalarized solutions around an isolated static BH. Solid, dashes and dotdashed lines correspond to zero, one or two nodes solution respectively. The black lines ($\ell=0$) match with previous literatue results\til\cite{Silva:2017uqg}. Because of spherical symmetry, scalar perturbations of an isolated BH in EsGB are insensitive to the specific values of $m$. Thus, each curve correspond to a specific, single value of $\ell$, regardless of the value of $m$.}	
\label{fig:BH_spscal_n012}
\end{figure}
As a consistency check, we first integrate Eq.\til\eqref{eq:KG_BBH} for a single static BH ($Z_0=0$), searching for static bound states, as the ones found in\til\cite{Silva:2017uqg,Doneva:2017bvd}. This means that in the following we seek only for solutions with 
\begin{equation}
\omega=0\,.
\end{equation}
Considering the quadratic coupling function in Eq.\til\eqref{eq:coupling_function}, a comparison with the results in Ref.\til\cite{Silva:2017uqg} is straightforward. In Fig.\til\ref{fig:BH_spscal_n012} we show different scalar bound states that correspond to unstable solutions around Schwarzschild BHs for the first three scalarized solutions, for $\ell=0,1,2$. Not all the values of $\eta/M^2$ provide static scalar non-trivial solutions. In fact, these bound states correspond only to a specific set of $\eta/M^2$. The corresponding values of the coupling parameter are summarized in table\til\ref{table:eta_values}.
\begin{table}[th] 
	\begin{tabular}{c||c}
		\hline
		\hline &
	\multicolumn{1}{c}{$\left(\eta/M^2\right)^{n\ell m}_{Z_0=0}$}\\
		\hline
		$\ell$ &  n=0\,\;\;\;\;\; n=1 \;\;\,\, n=2 \\ 
		\hline
		\hline
		0 & $2.902\;\;\;\,\,\,   19.50\;\;\;\,\,     50.93$\\
		1 & $8.282\,\,\;\;\;\,\,     29.82\;\;\;\,\,   65.84$\\
		2 & $16.30\,\,\;\;\;\,\,  42.97\;\,\,   \;\;  83.82$\\
		\hline
		\hline
	\end{tabular} 
	\caption{Values of the coupling constant $\eta$ corresponding to the static scalar bound states solutions around isolated BHs. Each value of $\eta/M^2$ refers to a different curve in Fig.\til\ref{fig:BH_spscal_n012}. The values for $\ell=0$ agree with the literature\til\cite{Silva:2017uqg}.}
	\label{table:eta_values}
\end{table}
Compare to previous literature\til\cite{Silva:2017uqg}, we evaluate the static unstable bound states also for $\ell>0$. These solutions will serve as benchmarks for the bound states solution in the BBH case, as we shall see in the next paragraph. 

Each entry in table\til\ref{table:eta_values} corresponds to a parabola in a $\left(\eta,M\right)$ plane. Non-linear studies including the scalar field backreaction on the spacetime geometry showed how hairy BHs solutions, end points of the tachyonic scalar instability, belong only to an infinite set of narrow bands in the $\left(\eta,M\right)$ plane\til\cite{Silva:2017uqg}. The values in table\til\ref{table:eta_values}, computed through a linear analysis, coincide only with one of the two ends of each band.

\noindent{\bf{\em Binary black hole spontaneous scalarization.}}
Let us turn now to the case of two BHs in a binary. Hence, we solve Eq.\til\eqref{eq:KG_BBH} for $Z_0\neq0$. As clear from the coefficients in Eq.\til\eqref{eq:q12lm}, scalar monopolar perturbations vanishes when $Z_0\neq 0$. Hence, the results obtained for isolated BHs hold when $\ell=0$. 

\begin{figure}
\includegraphics[width=8.5cm,keepaspectratio]{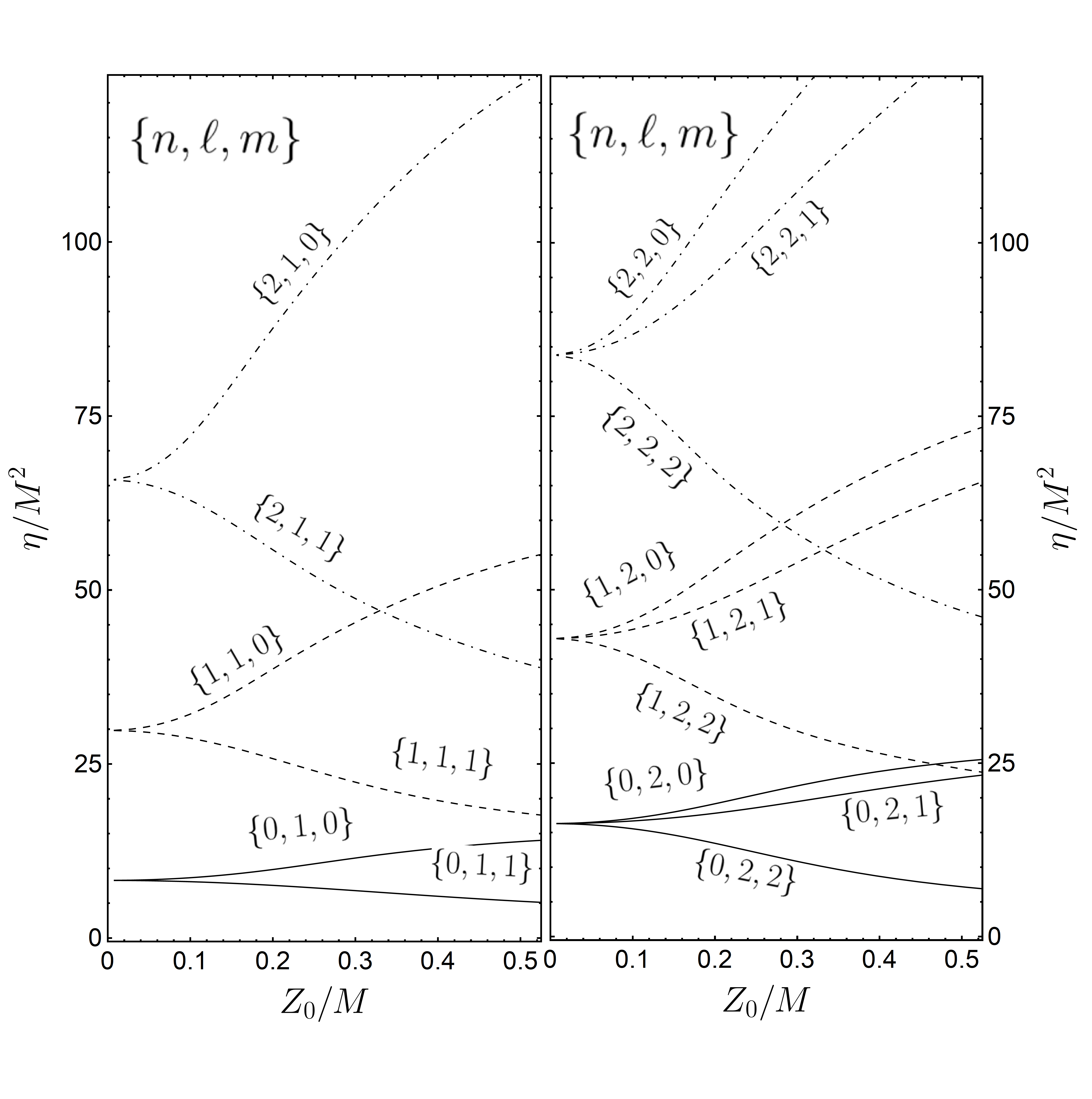}
\caption{Existence lines for the coupling constant of EsGB gravity, corresponding to static bound states solutions of Eq.\til\eqref{eq:KG_BBH}, as a function of the normalized geometrical BHs separation ($Z_0/M$). Each curve is labelled for different values of $\{n,\ell,m\}$. In both panels, the solid lines correspond to zero node solutions ($n=0$), the dashed to one node ($n=1$) and dotdashed to two nodes ($n=2$). Results for negative values of $m$ coincide with their positive $m$ counterpart, and therefore and not explicitly shown in the legend. All the different branches depart, respectively, from each value shown in Tab.\til\ref{table:eta_values}, previously evaluated for $Z_0=0$. {\bf Left panel}:   bound states associated with $\ell=1$. {\bf Right panel}: bound states associated with $\ell=2$.}	
\label{fig:BH_spscal_binary}
\end{figure}

For $\ell\geq 1$ instead, we compute how the specific values of $\eta/M^2$ shown in Tab.\til\ref{table:eta_values} vary as a function of the BHs separation. Results are summarized in Fig.\til\ref{fig:BH_spscal_binary}.
Different branches for the same $\ell$ refer to different values of the spherical 
harmonic index $m$. From Eq.\til\eqref{eq:q12lm} we may notice that each branch in Fig.\til\ref{fig:BH_spscal_binary} departs from the single BH value ($Z_0=0$) to larger values of $\eta/M^2$ if $q_1^{(\ell m)}>0, q_2^{(\ell m)}<0$, and to smaller ones if $q_1^{(\ell m)}<0, q_2^{(\ell m)}>0$. 
All the branches in Fig.\til\ref{fig:BH_spscal_binary} for which the value of $\eta/M^2$ decreases when $Z_0$ increases can be approximated by the following fit
\be \label{eq:fit_eta_Z0}
\frac{\eta}{M^2} \approx  \oo\frac{\eta}{M^2}\cc_{Z_0=0}^{n\ell m}-a^{n\ell m}\oo\frac{Z_0}{M}\cc^{3/2}\,,
\ee
accurate within $1\%$ for $0\leq Z_0/M\leq 0.4$. In the above fit, the first term on the r.h.s corresponds to each specific entry in Tab.\til\ref{table:eta_values} and $a^{n\ell m}$ is a constant that depends on the the number of nodes and on the angular indices. As an example, some of its values are $a^{011}=8.74,a^{022}=31.64,$ etc..

Finally, given the assumptions made to build the binary spacetime in paragraph {\bf{\em III}}, we stress that the results summarized in Fig.\til\ref{fig:BH_spscal_binary}, obtained for stationary backgrounds, have to be intended only as an indication of what happens to scalar fields in BBHs geometries, even when the BHs are left free to collide.

\noindent{\bf{\em V. Conclusions.}}
As depicted in Fig.\til\ref{fig:BH_spscal_binary}, BBH spacetimes in EsGB might suffer scalar fields instabilities. These results indicate that this process can happen {\it before} the final object is formed. On top of it, this unstable mechanism can be enhanced by BBH spacetimes, for smaller values of the coupling constant compare to the corresponding isolated BH case. In fact, the scalar field might grow significantly during the collision of BHs, whose masses would not allow to form a final black hole with a compactness sufficient to scalarized on its own. Such results remark once more the fundamental role that the strong field regime possesses during BHs collisions and coalescences: in order to perform consistent tests of alternative theories, we need waveforms that properly accounts for backreacting effects when high spacetime curvatures are involved. Notably, this work is a first step towards the study of the GWs produced by merging BHs in EsGB through the CLAP formalism.

\noindent{\bf{\em Acknowledgements.}}
%
I am indebted to Vitor Cardoso and Leonardo Gualtieri for important feedback and comments on this manuscript.

I also acknowledge partial financial support provided under the European Union's H2020 ERC Consolidator Grant ``Matter and strong-field gravity: New frontiers in Einstein's theory'' grant agreement no. MaGRaTh--646597, the European Union's Horizon 2020 research and innovation programme under the Marie Sklodowska-Curie grant agreement No 690904 and the GWverse COST Action 
CA16104, ``Black holes, gravitational waves and fundamental physics.''
Finally, I also acknowledge the Funda\c{c}ao para a Ci\^{e}ncia e a Tecnologia for financial support through Project~No.~UIDB/00099/2020, PTDC/MAT-APL/30043/2017 and the PD/BD/128232/2016 awarded in the framework of the Doctoral Programme IDPASC-Portugal.
%

\appendix

\section{Separating KG equation}\label{app:separateKG}
In this section we show how to separate the KG equation in EsGB with quadratic coupling function (see Eq.\til\eqref{eq:coupling_function}), 
\be \label{eq:KG_equation_appendix}
\square\Phi=-\frac{\eta}{4} \Phi\rgb\,.
\ee

Thanks to the CLAP of BBH, we start splitting both the box operator and $\rgb$ in powers of the BH separation $Z_0$, as shown in Eq.\til\eqref{eq:box_expansion}. Hence, up to leading order, Eq.\til\eqref{eq:KG_equation_appendix} takes the form
\be \label{eq:KG_pert}
\oo \square^{(0)}+ Z_0^2 \square^{(1)} \cc \Phi = -\frac{\eta}{4}  \oo \rgb^{(0)}+ Z_0^2 \rgb^{(1)} \cc \Phi\,,
\ee
where each contribution on the right hand side can be computed through the BBH spacetime in Eq.\til\eqref{eq:pert_Schw_CLAP_GR_BL},
\begin{align}\label{eq:R0andR1}
\rgb^{(0)}&=\frac{48 M^2}{r^6}\,,\nn\\
\rgb^{(1)}&=-\frac{\alpha\left(\theta\right)}{M r^6} \Bigg(r \left(r (2 M-r) \frac{d^2 g}{dr^2}+(r-5 M) \frac{d g}{dr}\right)\nn\\
&+3 g (4 M-r)\Bigg)\,,
\end{align}
with $\alpha\left(\theta\right)=1+3 \cos (2 \theta )$.

Thus, we expand the scalar field in scalar spherical harmonics $Y^{\ell m}(\theta,\phi)$, as in
Eq.~\eqref{eq:Phi_ansatz}, where the harmonic functions 
normalization reads as $ \int d\Omega \oo Y^{\ell m}\cc^* Y^{\ell'm'}=\delta_{\ell \ell'}\delta_{m m'}$.
As shown in Ref.\til\cite{Annulli:2021dkw}, since Schwarzschild's spacetime ($Z_0=0$) is spherically symmetric, the spherical harmonics are
eigenfunctions of the KG operator on $g^{(0)}_{\mu\nu}$, while the first order KG operator ($\square^{(1)}$) couples harmonics with different $\ell$. This means that each solutions of the zero-th order problem contains only one definite value of the index $\ell$. The index $m$, instead, always factors out from the equation since the background is axisymmetric. Since the first order KG operator is proportional to $Z_0^2$, we assume,
\begin{align} \label{eq:Phi_ansatz_2}
\Phi&=\frac{\psi_{\ell\,m}\left( t,r\right) Y^{\ell\,m}\left(\theta,\phi\right)}{r}\nn\\
&+Z_0^2 \sum_{\ell'\neq\ell}\frac{\psi_{\ell'\, m}\left( t,r\right) Y^{\ell'\, m}\left(\theta,\phi\right)}{r}\,.
\end{align}
It is important to remark that with the ansatz in Eq.\til\eqref{eq:Phi_ansatz_2} we restrict to excitations with a single value of $\ell$. Perturbation mixing multiple values of $\ell$s simultaneously may lower even more the threshold of instability. We leave this investigation for future work.

Inserting Eq.\til\eqref{eq:Phi_ansatz_2} in Eq.\til\eqref{eq:KG_pert} we find (dropping the $\left(\theta,\phi\right)$ dependence in the spherical harmonics),
\begin{align} \label{eq:KG_pert_2}
& \square^{(0)} \left[\frac{\psi_{\ell\, m} Y^{\ell\,m} }{r}\right]+ Z_0^2  \square^{(1)} \left[ \frac{\psi_{\ell\,m} Y^{\ell\,m} }{r} \right]\nn\\
&+ Z_0^2 \sum_{\ell'\neq \ell} \square^{(0)} \left[ \frac{\psi_{\ell'\,m} Y^{\ell'\,m} }{r}\right] =-\frac{\eta}{4} \Bigg[ \frac{\psi_{\ell\,m} Y^{\ell\,m} }{r} \rgb^{(0)} \nn\\
&+ Z_0^2 \frac{\psi_{\ell m}  Y^{\ell\,m} }{r}  \rgb^{(1)} +Z_0^2 \sum_{\ell'\neq \ell}\frac{\psi_{\ell' m}  Y^{\ell'\,m}}{r}\rgb^{(0)}\Bigg]\,.
\end{align}
Projecting the above equation on the complete basis of spherical harmonics, the components $\psi_{\ell'\,m}$ with $\ell'\neq\ell$ vanish, because the spherical harmonics are eigenfunctions of $\square^{(0)}$. Thus, the only remaining $O(Z_0^2)$ term on the l.h.s of Eq.~\eqref{eq:KG_pert_2} can be written explicitly as,
\begin{align}
  & \square^{(1)} \left[ \frac{\psi_{\ell\,m} Y^{\ell\,m} \left(\theta,\phi\right)}{r} \right]=-\frac{\partial Y^{\ell\, m}}{\partial \theta}\frac{3 g \sin (2 \theta ) }{8 M^2 r^3}\psi_{\ell\,m}\nn\\
&  +Y^{\ell\, m}\Bigg(- r\left(-r^2 f dg/dr+4 M g\right)\frac{\partial \psi_{\ell\,m}}{\partial r}-2r^3 f g\frac{\partial^2 \psi_{\ell\,m}}{\partial r^2}\nn\\
&+\left(-r^2 f dg/dr+2 g (\ell (\ell+1) r+2 M)\right)\psi_{\ell\,m}\Bigg)\frac{\alpha\left(\theta\right)\,}{16 M^2 r^4} \,,
\label{eq:KGZ02}
\end{align}
and we remind that $f=1-2M/r$.
 
As expected, the projection on $Y^{\ell\,m}$ of the $O(0)$ term on the l.h.s in Eq.~\eqref{eq:KG_pert_2} provides the standard form of the KG equation in
Schwarzschild's spacetime, 
\begin{align} \label{eq:order0KGproj}
  &-\frac{1}{r f}\Bigg(\frac{\partial^2\psi_{\ell\,m}}{\partial t^2}-f^2\frac{\partial^2\psi_{\ell\, m}}{\partial r^2}
  -f\frac{df }{dr}\frac{\partial\psi_{\ell\, m}}{\partial r}\nn\\
&+f\frac{\ell (\ell+1) r+2 M }{r^3}\psi_{\ell\, m}\Bigg)\,.
\end{align}
Finally, projecting the full Eq.\til\eqref{eq:KG_pert} on $Y^{\ell\, m}$ using Eq.\til\eqref{eq:R0andR1} and the results in Eqs.\til\eqref{eq:order0KGproj}-\eqref{eq:KGZ02}, we obtain the desired
{\it decoupled equation}
\beq \label{eq:KG_notortoise_app}
&&\frac{\partial^2 \psi_{\ell m}}{\partial t^2}+\frac{\partial^2 \psi_{\ell m}}{\partial r^2} \oo U_0+Z_0^2\tilde{U}_0\cc+\frac{\partial \psi_{\ell m}}{\partial r}  \oo U_1+Z_0^2\tilde{U}_1\cc\nn\\
&&+\psi_{\ell m} \oo \oo W_0+\frac{\eta}{4}\tilde{W}_0\cc +Z_0^2 \oo W_1+ \frac{\eta}{4} \tilde{W}_1\cc\cc=0\,,\nn\\
&&
\eeq
with radial potentials given by,
\begin{align}
\label{eq:potentials_notortoise}
&U_0(r)=-f^2\,,\nn\\
&\tilde{U}_0(r)=\frac{\left( r-2 M \right) f q^{(1)}_{\ell\,m} g}{8M^2r}\,,\nn\\
&U_1(r)= -f \frac{df}{dr} \,,\nn\\
& \tilde{U}_1(r)= -\frac{q^{(1)}_{\ell\,m} (2 M-r) \left(4 M g(r)-f r^2 dg/dr\right)}{16 M^2 r^3}\,,\nn\\
&W_0(r)=f\frac{\ell (\ell+1) r+2 M }{r^3}\,,\nn\\
&\tilde{W}_0(r)=\ooq \frac{48   M^2 (2 M-r)}{r^7}\ccq\,,\nn\\
&W_1(r)=\frac{q^{(1)}_{\ell\,m}}{16 M^2 r^4} (f r^2 (r-2 M) dg/dr\nn\\
&+2 g (2 M-r) (l (l+1) r+2 M))+3q^{(2)}_{\ell\,m}\frac{(r-2M)g}{4M^2r^3}\,,\\
&\tilde{W}_1(r)= -\frac{2  M q^{(1)}_{\ell m} \oo 2M-r\cc}{r^7}\Bigg(3 \oo 4M-r\cc \Delta_1 \nn\\
&+ r\oo \oo r-5M \cc\frac{d \Delta_1}{dr}  + \oo 2M-r\cc r \frac{d^2 \Delta_1}{dr^2} \cc \Bigg) \,.
\end{align}
The coefficients in Eqs.\til\eqref{eq:potentials_notortoise} are defined as
\begin{align} \label{eq:q1lm}
q^{(1)}_{\ell\,m}  \equiv&\int d\Omega \left( Y^{\ell\,m}\right)^* Y^{\ell\,m} \alpha\left(\theta\right)
  \,,\\ \label{eq:q2lm}
  q^{(2)}_{\ell\,m}\equiv &\int d\Omega \sin \theta \cos \theta \left( Y^{\ell\,m}\right)^*
  \frac{d Y^{\ell\,m}}{d\theta}  \,.
\end{align}
Since
\begin{equation}
\label{eq:q12lm}
q^{(1)}_{00}=q^{(2)}_{00}=0\,,
\end{equation}
the $\ell=0$ equation is not affected by the $O(Z_0^2)$ corrections. Instead, for $0<\ell\le2$ one gets,
\begin{align} 
&  q^{(1)}_{1-1}=q^{(1)}_{11}=-\frac{4}{5},\,q^{(1)}_{10}=\frac{8}{5}\,,\nonumber\\
&  q^{(1)}_{2-2}=q^{(1)}_{22}=-\frac{8}{7},\,q^{(1)}_{2-1}=q^{(1)}_{21}=\frac{4}{7},\,q^{(1)}_{20}=\frac{8}{7}\,,\nonumber\\
& q^{(2)}_{1-1}=q^{(2)}_{11}=\frac{1}{5},\,q^{(2)}_{10}=-\frac{2}{5}\,,\nonumber\\
& q^{(2)}_{2-2}=q^{(2)}_{22}=\frac{2}{7},\,q^{(2)}_{2-1}=q^{(2)}_{21}=-\frac{1}{7},\,q^{(2)}_{20}=-\frac{2}{7}\,.
\end{align}

\bibliographystyle{apsrev4}

\bibliography{References}

\end{document}